\documentclass[twocolumn,showpacs,preprintnumbers,amsmath,amssymb]{revtex4}

\usepackage{bm}
\usepackage{epsfig}
\usepackage[usenames]{color}

\newcommand{\nn}{\nonumber}

\renewcommand{\vec}[1]{{\pmb{ #1}}}

\newcommand{\rav}[1]{\left\langle#1\right\rangle}
\newcommand{\av}[1]{\left\langle#1\right\rangle}

\begin{document}


\title{Mean field theory for driven domain walls in disordered environments}

\author{Friedmar Sch\"utze}
\affiliation{Institut f\"ur Theoretische Physik, Universit\"at zu
K\"oln, Z\"ulpicher Str. 77, D-50937 K\"oln, Germany}

\author{Thomas Nattermann}
\affiliation{Institut f\"ur Theoretische Physik, Universit\"at zu
K\"oln, Z\"ulpicher Str. 77, D-50937 K\"oln, Germany}

\date{\today}

\begin{abstract}
We study the mean field equation of motion for driven domain walls in
random media. We discuss the two cases of an external constant as well as an 
oscillating driving force. 
Our main focus lies on the critical dynamics close to the depinning
transition, which we study by analytical and numerical methods.
We find power-law scaling for the velocity as well as the hysteresis loop area.

\end{abstract}


\maketitle

\section{\label{sec:intro}%
Introduction%
}

The interaction of elastic systems, like charge density waves
\cite{Fisher:PRL83,GruenerCDW:RMP88}, Wigner crystals \cite{Andrei:PRL88}, domain walls
\cite{Nattermann:JPC83,Nattermann:PSS85,Nattermann:PRL88}, 
dislocation lines \cite{ioffe+jpc87} or magnetic flux lines
with disorder is a problem of great technological importance.

More than two decades ago, in a seminal work,
D.S. Fisher has studied the depinning of charge density waves (CDW) 
from randomly distributed pinning centres \cite{Fisher:PRL83,Fisher:PRB85}
by an external constant (dc) field $h$. 
He showed that the depinning transition is a dynamical 
critical phenomenon where the velocity close to the depinning threshold $h_p$
plays the role of an order parameter exhibiting a power law behaviour
$v\sim (h-h_p)^\beta$.
Within the mean-field approach, developed in \cite{Fisher:PRL83,Fisher:PRB85},
the exponent $\beta$ was found to be $\beta=3/2$.
In a subsequent work \cite{NarayanFisher:PRB92}, the treatment
was extended to incorporate fluctuation effects using a functional
renormalisation group (FRG) approach. The latter is necessary
since the correlator of the random pinning forces develops a cusp singularity 
on sufficiently large length scales. This results in a modified mean field
exponent $\beta=1$, to which corrections in $\epsilon=4-d$ were
calculated in \cite{NarayanFisher:PRB92}.

A closely related problem concerns
the behaviour of driven interfaces in random environments,
as for example domain walls in disordered ferroic systems
\cite{Feigelman:JETP83,Nattermann:JPC85} or interfaces
between immiscible fluids that are pushed through porous media
\cite{rubio+prl89,Martys:PRL91}.
Fe{\u\i}gel'man \cite{Feigelman:JETP83} has considered the velocity 
corrections due to the disorder and estimated from that the depinning 
threshold. However, taken self-consistently, the theory
does not yield a finite depinning threshold.
Koplik and Levine \cite{KL:PRB85} used perturbation theory for
the associated mean-field model, which they derived analogous to
the procedure by Fisher for CDWs \cite{Fisher:PRL83}.
They found that the interface either follows a solution which moves with
constant velocity or it remains pinned. They were, however, not
able to extend their findings to the problem of spatially extended
interfaces, because their perturbative approach lacks
the necessary FRG analysis.
In a subsequent work, Leschhorn studied the mean field theory
for domain walls in a model which treats the disorder in a simplified
manner \cite{Leschorn:JPA92}. He considered a discretised lattice system
and allowed the random force field to take values out of three
possibilities only: $-1$, $0$ or $+1$. For his model, he also found
pinned and sliding solutions and determined the
velocity exponent as $\beta=1$, which is the same as for
CDWs when the disorder force has discontinuous jumps 
\cite{NarayanFisher:PRB92}.
Vannimenus and Derrida \cite{Vannimenus:JSP01} simplified the
Leschhorn model even further and were able to derive an
exact solution. The basic simplification of their model concerns 
the assumption of unit moves.
This means, that per unit time step a segment of the interface
either remains at rest, if the total force is smaller or equal
to zero. Otherwise, it moves exactly one step forward, 
independent of the magnitude of the force. 
Though this assumption admits an exact solution, 
the restriction to unit moves entails a non-uniform periodic 
behaviour of the mean velocity 
close to the threshold.
The time averaged velocity (over one period) has then
a different exponent $\beta=1/2$.

In 1992, the depinning transition of interfaces in random media
has been analysed within an FRG calculation, starting from an expansion
around a flat interface \cite{NSTL:JP2F,NSTL:APL}. 
Similar to the charge density wave
case, a cusp singularity of the force correlator develops giving rise
to a finite depinning threshold as well as a non-trivial exponent
$\beta=1-\epsilon/9$, where $d=4-\epsilon$ is the
interface dimension.
Later, this FRG calculation as been refined to include
also two-loop terms \cite{CDW:PRL01,CDW:PRB02}.

The FRG flow equations for dc driven interfaces have then been
extended to the case of an oscillating (ac) driving force \cite{GNP:PRL03}.
Combined with a scaling analysis it was possible to work out
the main characteristics of the velocity
hysteresis loop $v(h)$. In the limit of small frequencies,
scaling behaviour has been found.
The exponents of the remanent velocity at depinning
as a function of frequency have been determined for 
all dimensions $d<4$ and the results 
agree very well with the values obtained from a 
numerical study \cite{GNP:PRL03}.
The problem of ac-driven interfaces is also experimentally relevant.
Of special interest is the ac susceptibility of
ferroic systems \cite{Kleemann:PRL07,Kleemann:ARMR07} which 
gets a considerable contribution from the domain wall motion
\cite{LykNP:PRB99,NPV:PRL01}.
A phenomenological understanding of different regimes
has been reported in Ref. \cite{FedMS:PRB04}, where the concept of
waiting time distributions has been used.
Moreover, the perturbation theory for ac-driven interfaces 
in random environments has been examined \cite{Schuetze:PRE10}.
Further study of ac-driven elastic systems in disordered
media has been devoted to
vortex lattices \cite{Daroca:PRB10} and structural defects in liquid
crystals \cite{Jezewski:PRB08}.

The aim of this work is to study the properties of the mean field theory
for driven domain walls in random environments close to the depinning
threshold for dc- as well as for ac-driving forces.
Especially for the regime of small driving frequencies
the critical behaviour has not yet been investigated. 
We find scaling behaviour for small frequencies.
We separately discuss the situation of smoothly correlated randomness,
which is the usual starting point for any kind of FRG study of elastic
systems and a type of disorder the correlator of which has a cusp 
singularity at the origin, reflecting the fixed point solution of the 
FRG flow \cite{Fisher:PRL86,NSTL:JP2F}.

The present article is organised in the following way.  
After a brief derivation of the mean field equation of motion in section
\ref{sec:eom}, we focus on the special case of an adiabatic driving force
in section \ref{sec:dc}.
We generalize the arguments of Fisher \cite{Fisher:PRB85}
to the problem of non-periodic systems, in order
to find the scaling properties close to the depinning transition for
adiabatic driving.
Our analytical findings are supported by numerical analyses.
In section \ref{sec:ac}, we pay special attention to the small frequency scaling
of several quantities characterising the velocity hyteresis loop.
In the case of ac-drivings, analytical treatments are difficult, so we 
extensively have to resort to numerical investigations.
An outline about the numerical methods employed is provided in appendix
\ref{app:num}.

\section{\label{sec:eom}%
Equation of motion%
}
The equation of motion of a driven interface in a $d+1$-dimensional
inhomogeneous medium can be written as \cite{Feigelman:JETP83}
\begin{equation}
\label{eq:mean_field0}
\frac{ 1}{\gamma}\frac{\partial z}{\partial t}=
\Gamma_0{\nabla}^2 z+h(t)+  u_0g({\vec x},z),
\end{equation}
where $z({\vec x},t)$ denotes the height of the interface profile. 
The three terms on the right hand side of  equation (\ref{eq:mean_field0}) 
denote the curvature force, the  external driving force 
$h(t)=h_0\cos(\omega_0 t)$ and the random force $u_0g({\vec x},z)$ 
arising from the inhomogeneities of the medium. 
The coefficient $\gamma$ denotes the bare mobility.
In the present paper we consider a 
mean field version of this  equation.
Mean field theory is suitable to describe systems above the upper 
critical dimension.
Below this critical dimension, the mean field analysis with a disorder
correlator that has a cusp singularity at the origin (see below) 
yields the zeroth order
for the critical exponents of the depinning transition in an
$\epsilon$-expansion ($\epsilon=D_c-D$) \cite{NarayanFisher:PRB93}.
In App. \ref{app:realisation} we discuss 
the applicability of mean field theory in more detail.
To obtain the mean field equation, we replace 
the local curvature term by some 
long range interaction \cite{Fisher:PRB85,KL:PRB85}
\begin{equation}
{\nabla}^2z({\vec x})\equiv \sum_{\langle{\vec x},{\vec y}\rangle}
{z({\vec y})-z({\vec x})\over a^2}
\to{2d\over a^2}\left(\overline {z(t)}-{z({\vec x}})\right),
\end{equation}
where ${\vec y}$ and ${\vec x}$ are nearest neighbour lattice sites, $a$ is
the lattice constant and
\begin{equation}
\label{MFA}
\overline {z(t)}=\left({a\over L}\right)^d
\sum\limits_{{\vec x}}z({\vec x})\to L^{-d}\int d^d\vec x\,z({\vec x},t).
\end{equation}
in the continuum limit.
This replacement results in an effective equation for $z({\vec x},t)$
\begin{equation}
\frac{ 1}{\gamma}\frac{\partial z({\vec x},t)}{\partial t}=
{2d\over a^2}
\Gamma_0\left(\overline{z(t)}-z({\vec x},t)\right)+h(t)+ u_0g(z({\vec x},t),{\vec x}),
\label{eq:mean_field1}
\end{equation}
where $\vec x$ plays now the role of a label for $z(t)$ and $g(z)$. 
In the thermodynamic limit, $L\to\infty$, $\overline{z(t)}$ does not 
fluctuate and hence (\ref{eq:mean_field1})
can be solved for a given $\overline{z(t)}$ with the condition (\ref{MFA}). 
For each $\vec x$ the solution depends on the random force field configuration 
$\{g(z)\}$ for this particular value of $\vec x$. The average  over all positions, 
$L^{-d}\int d^d\vec x$ can finally be replaced  by the average over the 
different random force configurations $\int Dg(z)P(g(z))$, i.e.
\begin{equation}
\label{eq:gzmittel}
\overline{z(t)}\to \int Dg(z)P(g(z)) z(t)\equiv\rav{z(t)}.
\end{equation}
For the random force we assume  a Gaussian distribution with $\rav{ g(z)}=0$, and 
\begin{equation}
\label{eq:diskorrdef}
\rav{g(z)\,  g(z^{\prime})}=\Delta_{\ell}(z-z^{\prime}),
\end{equation}
where $\Delta_{\ell}(0)=1$ and $\int\Delta_{\ell}(z)dz\sim\ell$. 
Here, $\ell$ denotes the correlation length of the force correlator, 
i.e. $\Delta_\ell(z)\ll 1$ for $|z|\gg \ell$.
In this paper we are going to consider two different types of
correlators. We distinguish between a correlator that is
smooth and a correlator that shows a cusp singularity at the origin.

In the following we measure $z$ in units of $\ell$ and $t$ in units of 
$\ell/\gamma u_0$, such that
\begin{equation}
\label{eq:eom}
\frac{\partial z}{\partial t}=\Gamma\left(\rav{z}-z\right)+{h}\cos \omega t+ g(z),
\end{equation}
where $\Gamma=2d\Gamma_0\ell/(u_0a^2)$, 
$\omega=\omega_0\ell/(\gamma u_0)$ and $h=h_0/u_0$. 
The function
$g(z)$ obeys the same relations as above with $\ell=1$, and therefore we will skip
the index $\ell$ at $\Delta_\ell(z)$ from now on.
Thus, the theory depends on the three parameters $\Gamma, h$ and $\omega$. 
The mean field equation of motion is similar to that for charge density waves 
considered by Fisher \cite{Fisher:PRL86} if $g(z)$ is replaced by $\cos(z-\beta)$ 
where $\beta$ is a random phase.

The physical picture of the mean-field equation of motion is, that 
segments of the interface now behave like individual particles, each of which moves
in a distinct configuration of the disorder. Every particle is coupled
to the disorder averaged position, which in turn is determined self-consistently.

\section{\label{sec:dc}%
Zero frequency limit%
}

In this section, we consider a special case of the equation of
motion (\ref{eq:eom}), for which the driving force is constant in time
\begin{equation}
\label{eq:dceom}
\frac{\partial z}{\partial t}=\Gamma
\left(\rav{z}-z\right)+{h}+g(z).
\end{equation}
At sufficiently large driving force $h$, the average particle position
$\av z$ will move with constant velocity $v\ge 0$. In this case,
Eq. (\ref{eq:dceom}) can be written as
\begin{align}
\label{eq:slideom}
\partial_tz&=\Gamma vt+h-\Gamma z+g(z)
\nonumber\\
&=g(z)-[\Gamma(z-vt)-h].
\end{align}
\begin{figure}[t]
\begin{center}
\includegraphics[width=.9\columnwidth]{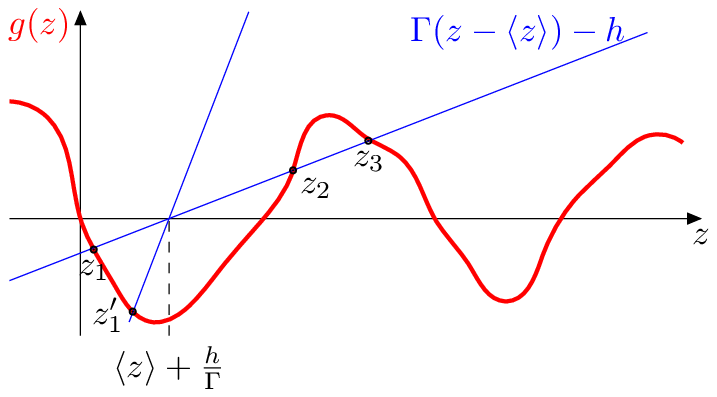} 
\caption{Plot of the left and the right hand side of equation 
(\ref{eq:static_sol}) for random forces with a smooth correlator.
For small $\Gamma$ there are several intersection points, for
small values of $\Gamma$ there is only one solution.
}
\label{fig:intersection1}
\end{center}
\end{figure}
In the following we will consider the case where the velocity is sufficiently
small $v\ll h$. The positions where $\partial_tz=0$ follow from the intersection 
of $g(z)$ with the straight line $\Gamma(z-vt)-h$ which moves to the right with 
velocity $v$ (cf. Fig. \ref{fig:intersection1}). 
For sufficiently small $\Gamma$ and smooth $g(z)$ there are in
general $2n+1$ intersections which we denote by $z_1<z_2<z_3<\ldots$. For $z<z_2$,
$z$ is driven towards $z_1$, for $z_2<z<z_4$ it is driven towards $z_3$ etc.
If the particle starts with an arbitrary initial value, it will first
develop towards the closest stable fixed point of (\ref{eq:slideom}),
where the particle velocity is small. Let us assume this is $z_1$. 
The force free point $z_1$ will then change according to 
$z_1(t)= vt+h/\Gamma+g(z_1)/\Gamma$. Eventually, the intersection
point $z_1(t)$ merges with $z_2(t)$ and then disappears. In this case
$z(t)$ will grow sufficiently fast until it reaches $z_3(t)$ and the
process repeats if we replace $z_n\to z_{n-2}$. Thus, the motion of the 
particle is jerky: periods of slow motion with velocity $v$ are 
intermitted by fast periods where the particle
is driven towards a new stable fixed point. Below, we will analyse this 
process in detail.

\subsection{%
The general picture%
}

A first overview results from considering some  limiting cases.

\begin{figure}[t]
\begin{center}
\includegraphics[width=\columnwidth]{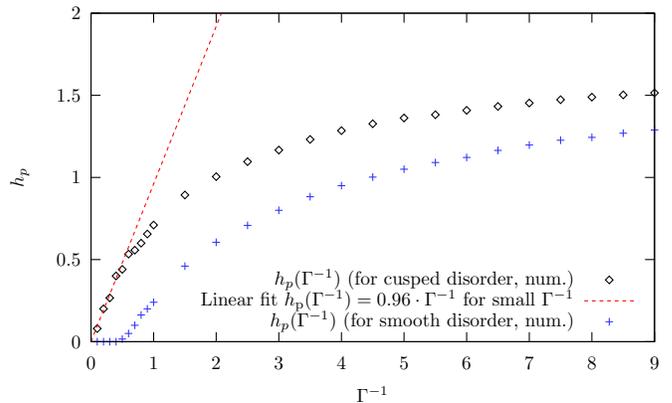} 
\caption{Depinning threshold as a function of  $\Gamma^{-1}$ in the case of a
dc-drive, $\omega=0$. For the case of a cusp-correlator  of the random forces 
(diamonds) the depinning threshold remains finite as long $\Gamma^{-1}$ is finite.  
For a smooth correlator  (crosses) the threshold vanishes for small $\Gamma^{-1}$ 
as expected from perturbation theory.}
\label{fig:hpgamma}
\end{center}
\end{figure}

(i) For large but finite $\Gamma$ we can apply perturbation theory.
To lowest non-trivial order one obtains for the mean
velocity (a short derivation of this expression can be found in
appendix \ref{app:pert})
\begin{equation}
\label{eq:vpertdc}
v=h+\int_0^{\infty}dt\,
e^{-\Gamma t} \Delta^{\prime}(vt)\,.
\end{equation}
The depinning
threshold $ h_{p,\pm}$ for $h\lessgtr 0$ follows from
taking the limit $v\to\pm 0$, $h\to\pm h_p\pm0$
\begin{equation}
\label{eq:hp_pert}
h_{p,\pm}\equiv -
\lim\limits_{\varepsilon\to
0}\Gamma^{-1}\Delta^{\prime}(\pm\varepsilon).
\end{equation}
Thus, the force
correlator has to have a cusp singularity to produce a finite
threshold.
If there is no cusp, perturbation theory in $\Gamma^{-1}$ signals 
the absence of a depinning threshold.  This argument applies however only to the 
region where perturbation theory is applicable, i.e. for $\Gamma\gg1$.
This perturbative result is in accordance with our numerical analysis,
as is shown in Fig. \ref{fig:hpgamma}.

As has been 
mentioned in the introduction, a cusp singularity in the correlator
emerges as a fixed point
solution of the functional renormalization group (FRG)
flow in $4-\epsilon$
dimensions and describes the effective randomness on scales
larger than the Larkin length. This leads to the existence
of a depinning threshold in all dimensions $d<4$. 
Of course, in the framework of
the mean field approximation an FRG study is senseless and a
correlator with a cusp singularity has to be included manually.
Nevertheless, as we already see here, in many aspects the
assumption of a correlator with a cusp gives different results 
compared to a smooth correlator. Incidentally, Narayan and Fisher
\cite{NarayanFisher:PRB92,NarayanFisher:PRB93} have used the mean
field solution for cusped disorder to expand around in order
to work out the critical behaviour of $4-\epsilon$-dimensional
systems.

(ii) Finally, we consider the case $\Gamma\ll1 $. For $\Gamma=0$ 
the equation of motion (\ref{eq:dceom}) can be integrated
\begin{equation}
\int_0^z\frac{dz'}{h+g(z')}=t.
\end{equation}
To calculate the integral we assume that $h>0$ and $h+g(0)>0$. Then for small $z$ 
the lefthand side is positive and hence $t$ as well, 
so the velocity is finite. However, since 
$g(z)$ is unbounded there is a value $z_1$ at which the denominator vanishes.
The integral is then dominated by the integration in the vicinity of $z_1$
Thus if $z$ approaches 
$z_1$ the time scale $t$ diverges and the velocity vanishes, 
the particle is pinned at $z=z_1$ . The same argument works for $h<0$.

\subsection{%
Static solution%
}

One special class of solutions to the equation of
motion (\ref{eq:dceom}) are the static solutions $z_s$ with 
$\partial_tz_s\equiv 0$. Here, we are going to analyse under which
circumstances such solutions can exist \cite{note1}.

From the equation of motion Eq. (\ref{eq:dceom}) 
it is clear that 
\begin{equation}
\label{eq:static_sol}
\Gamma\left(z_s-\langle z\rangle\right) - h=g(z_s)
\end{equation}
must be obeyed, i.e. the system has to be located at force-free positions.
\begin{figure}[t]
\begin{center}
\includegraphics[width=.9\columnwidth]{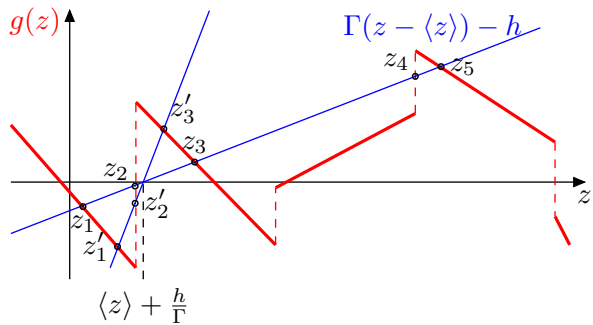}
\caption{Plot of the left and the right hand side of equation 
(\ref{eq:static_sol}) for random forces accociated to a scalloped
potential, which shows a cusp singularity in the correlator.
Random force realisations with a jump close to the
origin yield more than one solution even for very small values
of $\Gamma$.
}
\label{fig:intersection2}
\end{center}
\end{figure}
Besides Eq. (\ref{eq:static_sol}), one has to take into account that the 
self-consistency condition
\begin{equation}
\label{eq:static_sc}
h=-\langle g(z_s)\rangle,
\end{equation}
which follows from averaging (\ref{eq:static_sol}) holds.
The maximal value on the righthand side of (\ref{eq:static_sc})
is realised, if $z_s\equiv z_1$. Thus,
\begin{align}
\label{eq:hp}
h_p=-\av{g(z_1)}
\end{align}
is a critical field strength, above which no static solutions
are possible. Conversely, we can conclude that 
close to depinning all particles
are localised at the leftmost force free points.

Let us now apply this argument to the case  $\Gamma\gg1$. For a smooth potential 
as depicted in Fig. \ref{fig:intersection1} there is typically only one 
solution $z_1'$. For this single solution,
$g(z_1')$ can be positive or negative with equal probability. 
Thus, a pinned solution 
obeying $h=-\av{g(z_1')}$ does not exist apart from the case $h=0$. 
Hence, in this case the interface is never pinned, in agreement with our result 
from perturbation theory. The situation is different in the case when the random 
force exhibits infinite slopes as is shown in Fig. \ref{fig:intersection2}.
Then, due to the discontinuities there are in general
several solutions $z_i$ for any value of $\Gamma$
from which the leftmost ones dominate the behavior in the neighborhood of the 
depinning threshold. Of course, the larger the value of $\Gamma$ the smaller
is the fraction of disorder realisations which allow for more than one force
free solution. Thus, for large $\Gamma$, the depinning threshold is diminished
but finite.
The presence of infinite slopes is a special feature of disorder
forces, the correlator of which has a cusp singularity at the
origin. A detailed analysis of the cusped disorder, as is sketched in
Fig. \ref{fig:intersection2} is presented in
App. \ref{app:cuspcorr}, where we discuss how such a class of disorder
forces can be realised and derive the correlator explicitly.

To test these predictions, we have solved equation (\ref{eq:dceom})
numerically. 
The depinning threshold $h_p$ is plotted in 
Fig. \ref{fig:hpgamma} as a function of $\Gamma$. It is clearly seen that the 
threshold increases with $\Gamma^{-1}$, it vanishes for $\Gamma>\Gamma_c$ for smooth 
random force correlations.  For cusp correlations $\Gamma_c\to \infty$. 
These  findings support the results from perturbation theory.

\subsection{\label{subsec:dc:scal}%
Scaling behaviour above depinning%
}

Now, we consider the behaviour slightly above the 
depinning threshold $h\gtrsim h_p$, when 
$\av{z}=vt$ but $v\ll 1$.  
Our goal is to work out the scaling exponent $\beta$ for the
sliding velocity $v$, which we anticipate to vanish as a power law
\begin{align}
v\sim(h-h_p)^\beta.
\end{align}
To this aim, we solve the equation of motion in an approximate manner.
As the velocity of the interface is small, $v\ll 1$, we can also
expect that $\partial_tz\ll 1$ for most of the time.
Thus, $z(t)$ follows essentially from the vanishing of the righthand side 
of (\ref{eq:slideom}), which means that $z(t)$ stays close to the
leftmost fixed point $z_1(t)$.
Since the disorder averaged position $\av{z}$ is in motion,
we have to keep in mind, that the root of the straight line
in Figs. \ref{fig:intersection1} and \ref{fig:intersection2}
is now moving relative to $g(z)$.
The intersection point $z_1(t)$ satisfies the relation
\begin{equation}
\label{eq:z1defgl}
z_1(t)=vt+\Gamma^{-1}(h+g(z_1(t))).
\end{equation}
Without loss of generality, we restrict ourselves to $v>0$, 
so $z_1(t)$ moves now to the right. In this part of the motion, 
$z_1$ changes slowly (of order $v$).  Eventually, $z_1$ merges 
with $z_2$. Let us assume that this happens at $t=0$.
For further reference we denote 
\begin{align}
z_0\equiv z_1(0-)=z_2(0-).
\end{align}
For $t>0$, these two solutions disappear and the intersection point $z_3(0-)$ 
becomes the new leftmost intersection point, i.e.  $z_3(0-)\to z_1(0+)$,
so effectively $z_1$ jumps instantaneously. 
Thus, at $t>0$, the position $z(t)$ is not any more close to
a force free position and therefore it moves faster to approach
the new intersection point $z_1(t)$. The idea is now, that the mean
velocity $v$ is mainly determined by those disorder realisations,
which move fast. In order to determine the scaling exponent $\beta$ of $v$,
it is thus our task to work out a quantitative description of the motion 
of a particle in a certain 
disorder realisation during a period of time between two collapses of
force-free points.
The temporal distance between two jumps of the leftmost 
force-free position is approximately given by
\begin{align}
T= v^{-1},
\end{align}
because this is the time needed to travel through a correlated region
of the disorder (which is of length 1).
We denote the distance to the new leftmost intersection point $z_1(t)$
by
\begin{align}
\label{eq:thetadef}
\theta(t)=z(t)-z_1(t).
\end{align}
Note, that by definition $\theta(t)$ is negative.
Now, Eq. (\ref{eq:z1defgl}) yields the identity
\begin{align}
0&=\langle z(t)-vt\rangle
={1\over T}\int\limits_0^Td t\>\langle z_1(t)+\theta(t)-vt\rangle
\nonumber\\
\label{eq:scvh}
&={1\over T}\int\limits_0^Td t\>\langle\Gamma^{-1}h+\Gamma^{-1}g(z_1)+
\theta(t)\rangle.
\end{align}
Using Eq. (\ref{eq:hp}), we obtain from (\ref{eq:scvh})
\begin{align}
\label{eq:scalidintegral}
{h-h_p\over\Gamma}=-{1\over T}\int\limits_0^Td t\>\av{\theta(t)}.
\end{align}
The integral on the righthand side of (\ref{eq:scalidintegral})
depends on the velocity. But, in order to use
Eq. (\ref{eq:scalidintegral}) to determine the scaling exponent,
we have to describe the interface motion for $t>0$, i.e. in the region
of the fast motion between the previous and the new force free position.
We are going to do this separately for the two types of disorder
considered in this paper.

\subsubsection{%
Disorder with a smooth correlator
}

The motion of the interface position after the collapse of the
two leftmost force-free points is best analysed in several steps.
First of all, we note that at $t=0$ when $z_1$ and $z_2$ merge,
the relation
\begin{align}
\label{eq:merge_condition}
\Gamma=g'(z_0)
\end{align}
holds.
For $t\gtrsim 0$,
we can expect that $z(t)$ is still close to $z_0$, so we
can expand (\ref{eq:slideom}) around $z_0$. Writing
\begin{align}
\label{eq:zetadef}
\zeta(t)=z(t)-z_0
\end{align}
and using (\ref{eq:merge_condition}), we obtain
\begin{align}
\label{eq:zeta}
\partial_t\zeta=\Gamma vt+{g''(z_0)\over 2}\zeta^2+ O(\zeta^3),\quad \zeta(0)=0.
\end{align}
For small $\zeta$, we can neglect the second term on the righthand side 
and obtain
\begin{align}
\zeta(t)\approx {\Gamma vt^2/ 2}.
\end{align}
On time scales $t\ge t_0=2[\Gamma vg''(z_0)]^{-{1\over 3}}$ the second term on 
the righthand side of (\ref{eq:zeta}) dominates the time evolution and we obtain 
\begin{align}
\zeta(t)&\approx
{2\over g''(z_0)(t_d-t)},\quad 
t_d={3\over 2}t_0.
\end{align}
Clearly, this result can only be used until a time 
\begin{align}
t_1\simeq t_d-{2\over g''(z_0)},
\end{align}
for which $\zeta(t_1)\lesssim 1$ since we made an 
expansion in $\zeta$. 
It shows, however, that 
for $t_0\lesssim t\lesssim t_1$ the coordinate $z$ increases 
rapidly until it comes close to the new 
leftmost minimum $z_1(t)$.
For $t>t_1$, $\theta(t)$ is already close to zero
and therefore gives only higher order contributions
to the righthand side of Eq. (\ref{eq:scalidintegral}).
\begin{figure}[t]
\begin{center}
\includegraphics[width=.9\columnwidth]{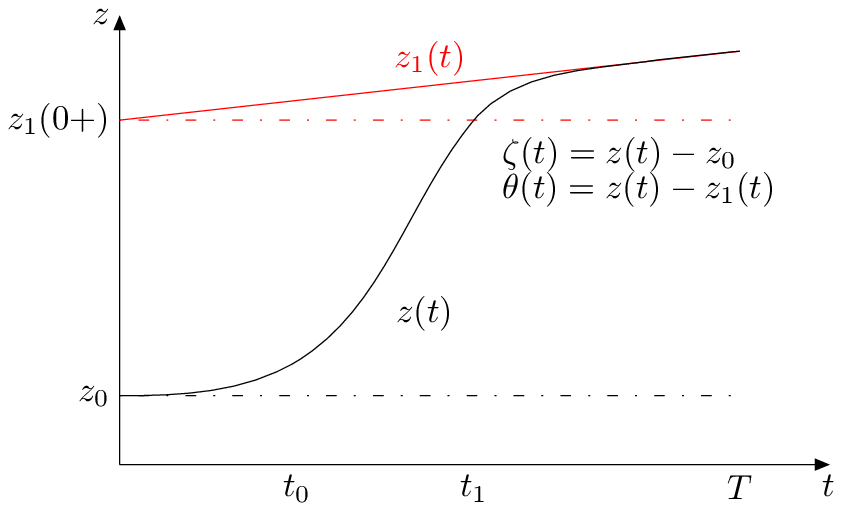}
\caption{Illustration of the motion $z(t)$ in between
two jumps in the case of smooth disorder.
}
\label{fig:smotion}
\end{center}
\end{figure}
The motion in between two jumps is sketched in Fig.
\ref{fig:smotion}

Now, we are going to evaluate the integral over 
$\theta(t)$ that occurs in Eq. (\ref{eq:scalidintegral}).
The equations (\ref{eq:thetadef}) and (\ref{eq:zetadef})
relate $\theta(t)$ and $\zeta(t)$
\begin{align}
\theta(t)=\zeta(t)+z_0-z_1(t).
\end{align}
The time dependence of $z_1(t)$ can be estimated 
from Eq. (\ref{eq:z1defgl}) as
\begin{align}
\partial_tz_1(t)&=v+\Gamma^{-1}g'(z_1(t))\partial_tz_1(t)
\nn\\
\Rightarrow\>
\partial_tz_1(t)&={\Gamma v\over\Gamma-g'(z_1(t))}={\Gamma v\over\Upsilon}+O(v^2),
\end{align}
where we have introduced $\Upsilon=[\Gamma-g'(z_1(0+))]$ for notational
convenience. Since $z_1(0+)$ is a stable fixed point, we have $\Upsilon>0$.
Using
\begin{align}
z_0-z_1(t)\simeq\theta(0)-{\Gamma v\over\Upsilon}t,
\end{align}
we obtain
\begin{align}
\int\limits_0^{t_0}dt\,\theta(t)=&\int\limits_0^{t_0}dt\,\left[\zeta(t)
+\theta(0)-{\Gamma v\over\Upsilon}t\right]
\nn\\
=&{\Gamma vt_0^3\over 6}+t_0\theta(0)-{\Gamma v\over\Upsilon}{t_0^2\over 2}
\nn\\
\label{eq:int0t0}
=&2v^{-{1\over 3}}\theta(0)[\Gamma g''(z_0)]^{-{1\over 3}}+
O(1).
\end{align}
Further, for $t_0<t<t_1$, we have
\begin{align}
\int\limits_{t_0}^{t_1}dt\,\theta(t)
=&\int\limits_{t_0}^{t_1}dt\,\left[\zeta(t)
+\theta(0)-{\Gamma v\over\Upsilon}t\right]
\nn\\
=&{2\over g''(z_0)}\ln{t_d-t_1\over t_d-t_0}+
\theta(0)(t_1-t_0)-
\nn\\
&{\Gamma v\over\Upsilon}{t_1^2-t_0^2\over 2}
\nn\\
\label{eq:intt0t1}
=&v^{-{1\over 3}}\theta(0)[\Gamma g''(z_0)]^{-{1\over 3}}
+O(\ln v).
\end{align}
As we have already said, the integral over the remaining time 
interval $[t_1;T]$
contributes to $O(1)$ only.
Thus, up to orders $O(v\ln v)$, from (\ref{eq:int0t0}) 
and (\ref{eq:intt0t1})
the expression on the righthand side
of Eq. (\ref{eq:scalidintegral}) follows as
\begin{align}
\label{eq:sdintegral}
{1\over T}\int\limits_0^Tdt\>&\langle\theta(t)\rangle
\simeq -v^{2\over 3}3\av{|\theta(0)|[\Gamma g''(z_0)]^{-{1\over 3}}}.
\end{align}
From (\ref{eq:scalidintegral}) and (\ref{eq:sdintegral}), 
we obtain therefore
\begin{equation}
v\sim {(h-h_p)^{3/2}\over\Gamma},
\end{equation}
i.e. $\beta=3/2$.

\subsubsection{%
Disorder with a cusped correlator%
}

As we have mentioned before, if $\Delta(z)$ has a cusp singularity,
the typical disorder force realisation exhibits discontinuous jumps,
as is depicted in Fig. \ref{fig:intersection2}.
A moment reflection shows, that a merging of two force free solutions
$z_1$ and $z_2$ is only possible at such a discontinuity of the force field.
The requirement, that $z_1$ is a stable fixed point entails that such a
discontinuity is given by an upward jump in the force field.
For the calculation we have to distinguish several cases.
\begin{figure}[t]
\begin{center}
\includegraphics[width=.9\columnwidth]{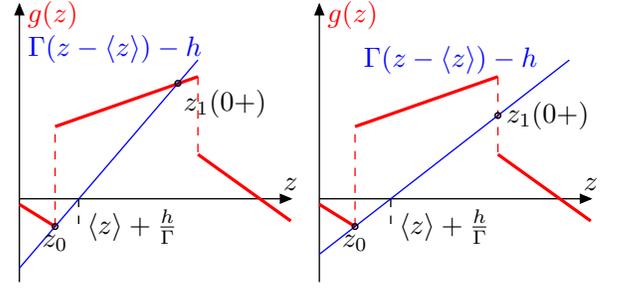} 
\caption{%
Left: This picture corresponds to our assumption for case 1,
that the new intersection point
$z_1(0+)$ is left of the next discontinuity of the disorder force $g(z)$.
It is obvious, that the inequality (\ref{eq:intersectioncondition}) 
has to be fulfilled.
Right: A scenario contrary to case 1 is possible. 
However, the basic fact
that the particle moves from the very beginning with a
velocity $g(z)-[\Gamma(z-\av z)-h]=O(1)$ 
and therefore needs a time $t_0=O(1)$ to approach $z_1(0+)$,
remains unchanged. So does the exponent $\beta$.
}
\label{fig:assumption}
\end{center}
\end{figure}%

\emph{Case 1:}\qquad
In this case we assume, that the next stable intersection point occurs before
the next discontinuity. Then, we have the inequality (cf. Fig. \ref{fig:assumption})
\begin{align}
\label{eq:intersectioncondition}
\Gamma>g'(z_0+).
\end{align}

It turns out that we have
to solve the equation of motion in two time regimes. First, close to $t=0$,
$z(t)$ is in the vicinity of $z_0$ and we consider again the equation for 
\begin{align}
\zeta(t)=z(t)-z_0.
\end{align}
Now, since the merging of two fixed points 
occurs at the discontinuities of the potential,
Eq. (\ref{eq:merge_condition}) is not meaningful, but instead $z_0$
fulfills the equation 
\begin{equation}
\label{eq:merge_condition_cusp}
\Gamma z_0=g(z_0-) +h.
\end{equation}
Using Eq. (\ref{eq:merge_condition_cusp}), it is easy to see
that the equation of motion for $\zeta(t)$ takes the form
\begin{align}\nonumber
\partial_t\zeta(t)&\approx\Gamma(vt-z_0)+g(z_0+)+
(g'(z_0+)-\Gamma)\zeta+h
\\
\label{eq:cuspzetaeom}
&=\Gamma vt-(\Gamma-g'(z_0+))\zeta(t)+\delta g.
\end{align}
Here, $\delta g=g(z_0+)-g(z_0-)$ denotes the jump of $g(z)$ which is of order one. 
Integration of (\ref{eq:cuspzetaeom}) gives for  short times $t\gtrsim 0$
\begin{equation}
\zeta(t)\approx {\delta}g\, t.
\end{equation}
This result is approximately correct for $t<t_0$ with
\begin{align}
t_0=(\Gamma-g'(z_0+))^{-1}.
\end{align}
Note, that due to (\ref{eq:intersectioncondition}) the time $t_0$ is 
always finite and positive, in fact generically of order $O(1)$.
For $t>t_0$, 
also the term in Eq. (\ref{eq:cuspzetaeom}) proportional to 
$\zeta(t)$ becomes relevant. Now,
$z_0+\zeta(t_0)$ has to be compared with $z_1(0+)$ 
which is the new leftmost intersection point for $t>0$. 
From (\ref{eq:z1defgl}) we deduce that $z_1(0+)$ fulfills the 
equation
\begin{equation}
\Gamma z_1(0+)\approx h+g(z_0+)+g'(z_0+)(z_1(0+)-z_0),
\end{equation}
from which we obtain
\begin{align}
z_1(0+)&\approx\frac{h+g(z_0+)-g'(z_0+)z_0}{\Gamma -g'(z_0+)}
\nonumber\\
&={[\Gamma-g'(z_0+)]z_0+\delta g\over\Gamma -g'(z_0+)}
=\zeta(t_0)+z_0=z(t_0).
\end{align}
In the second step, we have replaced $h$ using
Eq. (\ref{eq:merge_condition_cusp}).
Thus, after the time $t_0$ the particle has reached already the new 
intersection point $z_1$.

To determine the exponent $\beta$, we want to employ equation (\ref{eq:scalidintegral}) 
again. 
For $t\le t_0$, the relevant function $\theta(t)$ as has been obtained so far reads
\begin{align}
\theta(t)=z_0-z_1(t)+\zeta(t)\approx z_0-z_1(t)+\delta gt.
\end{align}
To approximate the time dependence of $z_1(t)$, we expand $z_1(t)$ around $z_1(0+)$ 
and get
\begin{align}
z_1(t)\simeq z_1(0+)+\dot z_1(0)t.
\end{align}
Here, $\dot z_1(0)$ can be deduced from the defining equation (\ref{eq:z1defgl}), 
it follows as
$\dot z_1(0)=\Gamma vt_1$ with 
\begin{align}
\label{eq:t1def}
t_1=(\Gamma-g'(z_1(0)))^{-1}.
\end{align}
Thus, in the regime where $\theta(t)$ changes fast, i.e. for $t\le t_0$, we can write
\begin{align}
\label{eq:thetatklt0}
\theta(t)\approx z_0-z_1(t)+\zeta(t)\simeq\delta g(t-t_0)-\Gamma vt_1\,t.
\end{align}
This shows, that for $t>t_0$, $\theta(t)=O(v)$.
However, the time scale $t_0$ is of the order $O(1)$, and is thus
small compared to $T$, $t_0\ll T$. Therefore, it is important to
carefully analyse the function $\theta(t)$ also for $t>t_0$.
For $t>t_0$ we expand around $z_1(t)$ and
the approximated equation of motion reads
\begin{align}
\partial_tz&\simeq\Gamma(vt-z)+g(z_1)+g'(z_1)(z-z_1)+h=
-{1\over t_1}\theta(t),
\end{align}
where $t_1$ is defined in
(\ref{eq:t1def}). Then, using (\ref{eq:z1defgl})
and (\ref{eq:thetadef}), we find
\begin{align}
\partial_t\theta&\simeq-{1\over t_1}\theta(t)-\Gamma vt_1.
\end{align}
The solution to this equation, matching with equation (\ref{eq:thetatklt0})
gives
\begin{align}
\label{eq:thetatgrt0}
\theta(t)\approx-\Gamma vt_1^2+\Gamma vt_1(t_1-t_0)\,e^{-(t-t_0)/t_1}.
\end{align}
\begin{figure}[t]
\begin{center}
\includegraphics[width=.9\columnwidth]{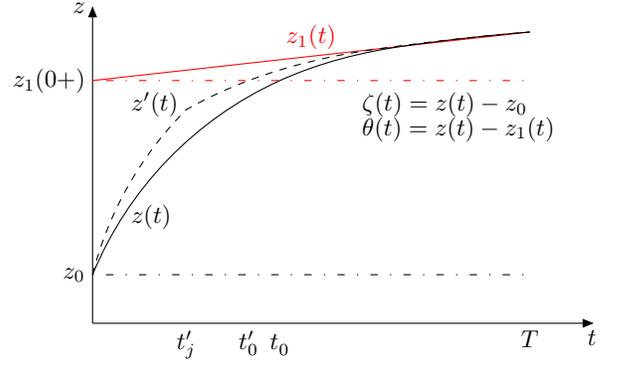}
\caption{Illustration of the motion $z(t)$ in between
two jumps in the case of disorder with a cusped correlator.
The difference between the absence (case 1, solid line) and the
presence (case 2, dashed line) of a discontinuity of the force
field $g(z)$ in between $z_0$ and $z_1(0+)$ is the appearance of
a sharp kink in the curve $z'(t)$ at $t=t'_j$.
}
\label{fig:cmotion}
\end{center}
\end{figure}
The motion $z(t)$ in between two jumps is sketched in Fig.
\ref{fig:cmotion}

Now, 
we can determine $\beta$ using equation (\ref{eq:scalidintegral}).
In calculating
\begin{align}
-{1\over T}\int\limits_0^Tdt\,\av{\theta(t)}
\simeq v\av{{\delta gt_0^2\over 2}+\Gamma t_1^2}+O(v^2),
\end{align}
we have decomposed the integral into the intervals $0\ldots t_0$ and $t_0\ldots T$,
respectively. This gives
\begin{align}
\label{eq:betacusp}
v&\sim{h-h_p\over\Gamma},
\end{align}
from which we conclude, that in the case of cusped disorder the
velocity exponent is $\beta=1$.

\emph{Case 2:}\qquad
Now, we have to discuss what can change if there is a discontinuity
of $g(z)$ in between $z_0$ and $z_1(0+)$.
One possible scenario for this case is depicted in the right part of
Fig. \ref{fig:assumption}. We are going to discuss now, that our main result
$\beta=1$ remains unchanged.
Indeed, as can be concluded from our previous calculation,
the essential point that lead to the exponent $\beta=1$ was the fact,
that $z(t)$ approaches $z_1(0+)$ on a time scale $t_0$ which is of order
$O(1)$. Responsible for this is, that immediately after a collapse
of the leftmost intersection point, the particle starts to move with a 
velocity of order $\delta g=O(1)$. This remains unchanged.
In Fig. \ref{fig:cmotion} we have also sketched the motion 
when a discontinuity occurs in between $z_0$ and $z_1(0+)$.
The respective quantities in Fig. \ref{fig:cmotion} carry a prime.
The only effect of the discontinuity that is crossed at a time $t'_j$
is a singularity of the velocity $\dot z'(t)$ at $t=t'_j$.
Therefore, the fundamental characteristics
of the motion remain unchanged. 
Thus, in case of $k$ jumps of $g(z)$ the foregoing calculation remains
basically unchanged, apart from the fact, that one should now
decompose the motion in more parts: $[0;t_{j_1}]$,
$[t_{j_1};t_{j_2}]$, \ldots, $[t_{j_k};t_0]$,
$[t_0;T]$. Of course, this consideration changes
the prefactor in Eq. (\ref{eq:betacusp}), which is, however, 
anyway beyond our accuracy.

\begin{figure}[t]
\begin{center}
\includegraphics[width=\columnwidth]{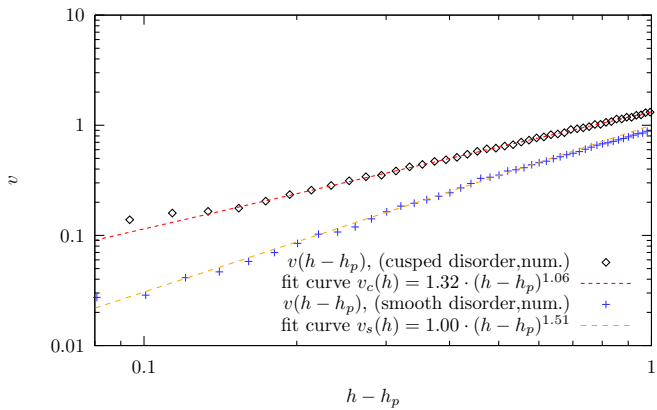} 
\caption{The velocity as a function of $h-h_p$ for $\Gamma=0.67$
in a double logarithmic plot. The numerically determined 
exponent for this measurement is $\beta=1.06\pm0.08$ for cusp-like singularity of 
$\Delta$ (diamonds) and $\beta=1.51\pm0.08$ for smooth force correlation (crosses).
}
\label{fig:beta}
\end{center}
\end{figure}%
The two exponents $\beta=3/2$ for smooth and $\beta=1$ for cusped disorder
are confirmed by our numerical solution as depicted in Fig. \ref{fig:beta}.

\section{\label{sec:ac}%
Finite Frequencies%
}

\begin{figure}
\includegraphics[width=\columnwidth]{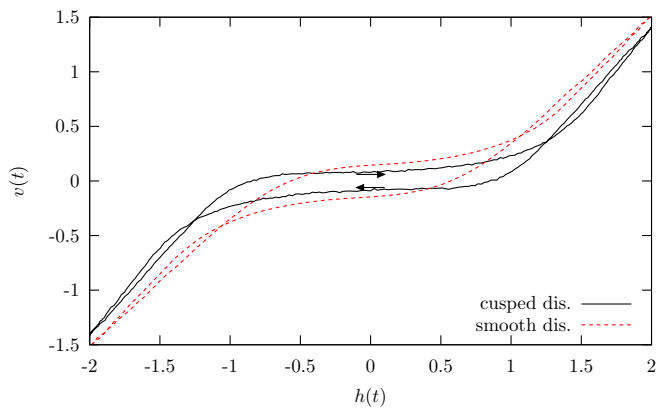}
\caption{In the presence of an ac driving field, a velocity hysteresis emerges.
In this picture we illustrate these hystereses for the cusped and the smooth
disorder for $\Gamma=0.5$, $h=2.0$ and $\omega=0.1$.
The inner hysteresis is traversed clockwise, the outer loops are passed
through counter-clockwise}
\label{fig:hystereses}
\end{figure}
In the finite frequency case, 
the disorder average over the solutions to the equation of motion
(\ref{eq:eom}) forms a hysteresis in the $v$-$h$-plane,
as is illustrated in Fig. \ref{fig:hystereses} for the two types of
disorder considered here. The hystereses are invariant under the transformation
$v\to -v$ and $h\to -h$. This can be explained directly using the equation
of motion (\ref{eq:eom}) and a statistical inversion symmetry. 
Taking the disorder average of (\ref{eq:eom})
yields
\begin{align}
\partial_t\av{z}=v=h\cos\omega t+\av{g(z(t))}.
\end{align} 
It is easy to see, that the aforementioned symmetry under $v\to -v$ and $h\to -h$ 
holds true if the probability density $P(g)$ (cf. Eq. (\ref{eq:gzmittel})) 
obeys $P(g)=P(\hat g)$ where $\hat g(z)=-g(-z)$. This is obviously the case for
our assumption of Gaussian disorder (cf. Sec. \ref{sec:eom}).

\subsection{\label{subsec:ac:qualitative}%
Qualitative discussion of the motion%
}

\begin{figure}
\includegraphics[width=.9\columnwidth]{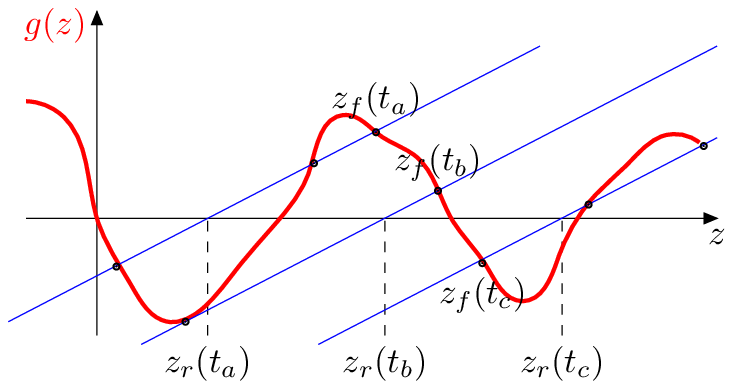}
\caption{%
At $h(t_a)\approx -h_p$, the particle is close to the
rightmost force free point
$z_f(t_a)$. This intersection point moves, due to the change of the zero
$z_r(t)=\av{z}(t)+h(t)/\Gamma$. The particle is following this
point. At a later time $t_c$, when
$h(t_c)\approx h_p$ this force free point has become the leftmost one.
}
\label{fig:acnegmot}
\end{figure}
To understand the shape of the hysteresis, we consider the motion of a particle
for half of a period for the case $h\gg h_p$ and small frequency $\omega\ll\Gamma$.
We start at a time $t=0$, when $h(0)=-h_p$ and the field increases.
Then, we can expect each particle to be located close to the
rightmost force free point, i.e. the rightmost solution of
\begin{equation}
\label{eq:acintersection}
\Gamma\left(z_f(t)-\av{z}(t)\right) - h(t)=g(z_f(t)).
\end{equation}
In Fig. \ref{fig:acnegmot} it is illustrated, that due to the change of 
the driving field towards larger values, the root of the straight
line, given by
\begin{align}
z_r\equiv\av{z}(t)+h(t)/\Gamma
\end{align}
moves with a velocity
\begin{align}
\label{eq:dotzr}
\dot z_r=v(t)+{\dot h(t)\over\Gamma}.
\end{align}
Since $\dot h(t)>0$, this velocity is positive although the value of the field
is still negative. Therefore, also the intersection point $z_f(t)$ 
to which the particle is connected, moves to the right.
This fact is observed in the hysteresis loop, illustrated
in Fig. \ref{fig:hystereses}.
\begin{figure}
\includegraphics[width=.9\columnwidth]{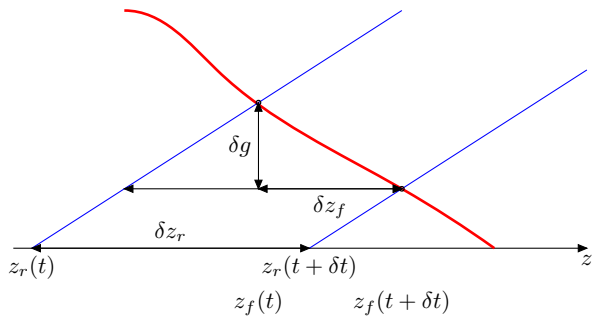}
\caption{%
Illustration for the velocity estimate
}
\label{fig:vest}
\end{figure}
Actually, this understanding allows to estimate the velocity in
simple geometrical terms. Using the notation explained in Fig. \ref{fig:vest},
we have ${\delta g/\delta z_f}\simeq g'(z_f)$ and thus
\begin{align}
\delta z_r-\delta z_f=-{\delta g\over\Gamma}=-{g'(z_f)\delta z_f\over\Gamma},
\end{align}
from which we conclude
\begin{align}
\delta z_r={\Gamma-g'(z_f)\over\Gamma}\delta z_f.
\end{align}
Now, Eq. (\ref{eq:dotzr}) yields
\begin{align}
{\delta z_r\over\delta t}={\Gamma-g'(z_f)\over\Gamma}{\delta z_f\over\delta t}
\simeq{\delta z_f\over\delta t}+{\dot h(t)\over\Gamma}.
\end{align}
Solving the last approximate equality for $\delta z_f/\delta t$, we obtain
\begin{align}
\label{eq:v_fest}
\dot z_f\simeq -\dot h/g'(z_f).
\end{align}

During the motion of $z_f(t)$, other intersection points to the
left of $z_f(t)$ vanish, and new solutions to the right emerge.
Finally, when $h(t_1)\approx h_p$, $z_f(t_1)$ has become the 
leftmost intersection point. From approximately this time
on it happens, that occasionally in some
disorder realisations $z_f(t)$ merges with an unstable fixed point
and vanishes, so that the particle moves fast in order to catch up
with the new leftmost force free point. This procedure has already been discussed
earlier in Sec. \ref{sec:dc}.
Since the velocity of a particle is given by the difference 
between $g(z)$ and
the straight line $\Gamma(z-\av{z})-h$, it must fall 
back behind the leftmost intersection 
point to speed up.
This can only happen due to the disappearance
of force free points. Thus, the velocity grows slowly
because after each jump the particle moves fast and
thus approaches again the new intersection point.
On the other hand, 
by virtue of Eq. (\ref{eq:dotzr}), the larger $v(t)$ 
the faster $z_r(t)$ and
thus also the faster the intersection points move.
This leads to a positive feedback and entails a strong slope
when the velocity is large enough such that 
the particle is no longer able to approach a force free point
before the next jump sets in.
Finally, far above $h_p$ the particle is depinned.
After the driving force has reached its maximum it decreases.
Note that the root of the straight line $z_r$ has now a
velocity smaller than $v(t)$, because $\dot h$ is negative.
Therefore, the particle position $z(t)$
approaches $z_r(t)$ and slows down.
Hence, $\dot h$ is a measure also for the decrease of $v(t)$.
On approaching $h_p$ from above, all
particles are still depinned and hence far enough behind the leftmost
intersection point, so that the latter has only little influence on the motion
of the particle and the velocity decays with the same slope all
the time. Only when $v(t)$ has passed below $\dot h/\Gamma$,
$z_r$ moves in the negative direction and thus the intersection points
as well. This means, that the leftmost intersection point
approaches the particle \emph{before} it is pinned. After the
particle is a little to the right of the leftmost force free position,
which happens about when $h(t)\approx h_p$, the velocity is negative. 
Now, the same procedure starts in the other direction.

As $\omega\to 0$, the hystereses approach
the depinning curve that has been discussed in the previous chapter.
In the following, we are going to take a closer look on the details of this
limiting process.

\subsection{%
Velocity exponents%
}

First, we want to work out, how 
\begin{align}
\label{eq:vh0def}
v_{h_0}\equiv |v(h=0)|
\end{align}
approaches zero as
$\omega\to 0$.
As we have explained in Sec. \ref{subsec:ac:qualitative},
the particle in each disorder realisation stays close to
a force-free point, that we have agreed to label $z_f(t)$.
The velocity $\partial_tz$ of the particle
is now determined solely by the velocity of the
force free position $z_f$ that
we are now going to calculate in a more accurate way than our
estimate from Eq. (\ref{eq:v_fest}).
Let $t_0$ be the point in time, at which $h(t_0)=0$. On time scales that are 
small compared to the period $\omega^{-1}$, we can linearly expand the driving
field around $t_0$
\begin{align}
\label{eq:htexp1}
h(t)\simeq -h\omega (t-t_0).
\end{align}
Further, we want to expand (\ref{eq:acintersection}) around $z_f(t_0)$. 
For small distances
in time we can neglect possible changes in the velocity and write
\begin{align}
z_f(t)\simeq z_f(t_0)+v_f(t-t_0),
\end{align}
where $v_f=\partial_tz_f$ is a shorthand notation.
Using (\ref{eq:htexp1}) as well as $\av{z}(t)\simeq \av{z}(t_0)-v_{h_0}(t-t_0)$, 
we have
\begin{align}
0=&\Gamma\big[z_f(t_0)+v_f(t-t_0)-\av{z}(t_0)+v_{h_0}(t-t_0)\big]+
\nn\\
&h\omega(t-t_0)-g(z_f(t_0))-g'(z_f(t_0))v_f(t-t_0)
\\
\label{eq:glfvh0}
=&(t-t_0)\big[\Gamma(v_f+v_{h_0})+h\omega-v_fg'(z_f(t_0))\big]
\end{align}
Since this should hold for small but finite $|t-t_0|$, the
expression in the rectangular brackets has to vanish.
Solving (\ref{eq:glfvh0}) for $v_f$, taking the disorder average and 
using the self-consistency condition $\av{v_f}=-v_{h_0}$ finally yields
\begin{align}
v_{h_0}=-{h\omega\over\Gamma-\av{[\Gamma-g'(z_f(t_0))]^{-1}}^{-1}}.
\end{align}
\begin{figure}
\includegraphics[width=\columnwidth]{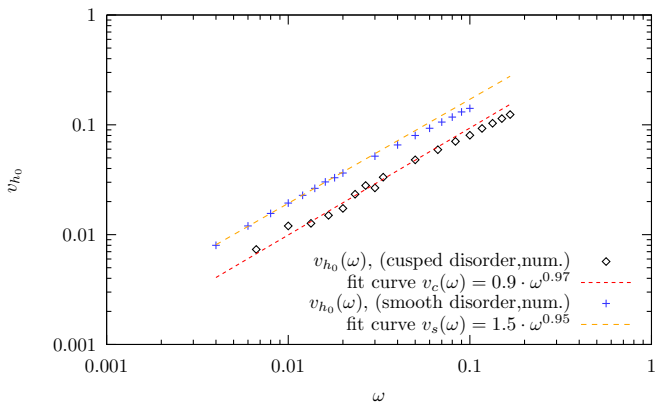}
\caption{The velocity $v_{h_0}$ as a function of frequency. The plotted
data correspond to numerical measurements at $\Gamma=0.33$ 
for cusped and $\Gamma=0.2$ for smooth disorder.
For both types of disorder, the exponents are close to 1
($\kappa_c=0.97\pm0.07$ and $\kappa_s=0.95\pm0.04$) in 
agreement with our analytical derivation given in the 
main text.}
\label{fig:vh0}
\end{figure}
Since $g'(z_f(t_0))<0$, which expresses the reasonable assumption that
$z_f$ is a stable force free position, $v_{h_0}$ is indeed positive, which
must be the case by its definition (\ref{eq:vh0def}).
Note, that our derivation so far does not make any assumption about the disorder
correlator, whence it holds for cusped as well as for smooth disorder.
In conclusion, for $\omega\to 0$ the width of the hysteresis at $h=0$
behaves as $v_{h_0}\sim\omega^\kappa$ with $\kappa=1$ for either kind of disorder.
This exponent is verified by our numerical analysis, cf. figure
\ref{fig:vh0}.

\begin{figure}
\includegraphics[width=\columnwidth]{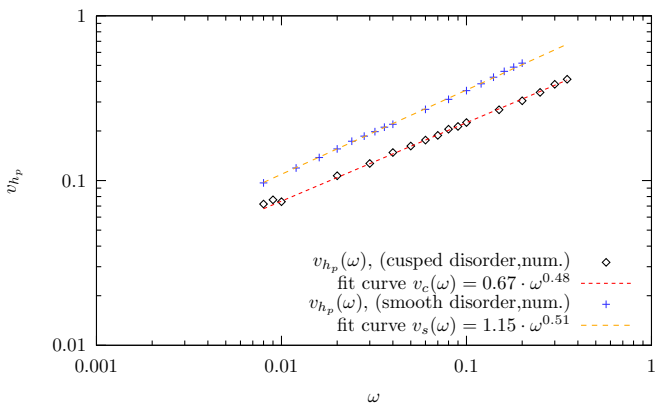}
\caption{The velocity $v_{h_p}$ as a function of frequency. The plotted
data correspond to numerical measurements at $\Gamma=0.33$ 
for cusped and $\Gamma=0.2$ for smooth disorder.
For both types of disorder, the exponents are close to $1/2$
($\mu_c=0.48\pm0.02$ and $\mu_s=0.51\pm0.01$).}
\label{fig:vhp}
\end{figure}

Another interesting quantity to look at is
\begin{align}
\label{eq:vhpdef}
v_{h_p}\equiv |v(h=h_p)|,
\end{align}
of which we want to work out the limiting behaviour for $\omega\to 0$.
As $h(t)$ increases further from $0$ towards $h_p$, more and more of the
force free points $z_f$, which the true positions in the disorder realisations are
following, become the rightmost ones, so that occasionally jumps occur.
On closely approaching $h(t)=h_p$ the dominant contribution originates from these
jumps, which severely affects the exponent, so that $v_{h_p}\sim\omega^\mu$ with
$\mu\simeq 1/2$, as can be inferred from our numerical analysis, shown in
figure \ref{fig:vhp}.
This exponent is again independent of the shape of the disorder correlator
at the origin (smooth or cusped).
An analytical derivation of this exponent is much more complicated than it was the
case for $\kappa$ and in fact we did not find a rigorous prediction.
For the finite dimensional case in $4-\epsilon$ dimensions, the
exponent $\mu$ has been found as $\mu=\beta/(\nu z)$, where $\nu$ denotes
the correlation length exponent and $z$ the dynamical exponent \cite{GNP:PRL03}.

\subsection{%
The area of the hysteresis loop%
}

Next, we want to investigate the limiting behaviour of the hysteresis area.
The physical meaning of the loop area can be concluded from the energy
balance of an overdamped system. For the change of the disorder averaged
potential energy in time, we find (cf. App. \ref{app:energy})
\begin{align}
\label{eq:potchange}
\partial_t\av E=h(t)v(t)-\av{(\partial_t z)^2}.
\end{align}
Here, $h(t)v(t)$ measures the energy gain through the work per unit time that is done 
by the external field and $\av{(\partial_t z)^2}$ measures the energy loss per unit time 
due to dissipation.
The area of the hysteresis loop is determined via
\begin{align}
\label{eq:ahystdef}
A_\text{hyst}(\omega)&=\oint v(t)\,dh=\int\limits_0^Tv(t)\dot h(t)\,dt.
\end{align}
This means, the loop area denotes the integrated change in work per unit time
due to the change of the external field.

Note, that in the case of a double hysteresis (which occurs for large $h$,
when the motion of the system over one period extends on average over more
than one valley of the disorder potential),
the area is given by the area of the 
inner hysteresis minus the area of the two outer hystereses 
(cf. Fig. \ref{fig:hystereses}). Formally, this is because
the inner hysteresis is traversed clockwise, whereas the outer loops are
passed through counter-clockwise.
Physically, this can be understood as follows.
Starting from $h=0$ at the branch of increasing $h(t)$, the external
field works against the potential gradient due to elastic energy and disorder.
On going over into the regime
of the outer loops, sliding behaviour sets in and thus the potential
energy, stored so far in the system, adds to the work done by the external field.
This fact is responsible for the steep slope at the beginning of the outer loop.
In other words, during the period in the outer loops, the external field does not
any more work against a potential gradient, but together with
the potential energy the system is accelerated. 

To work out the hysteresis loop area 
as $\omega\to 0$, we distinguish three cases. 

\begin{figure}
\includegraphics[width=\columnwidth]{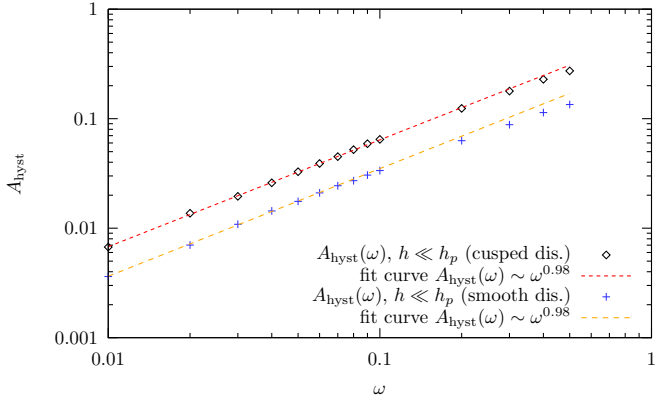}
\caption{The area of the hysteresis loop is plotted as a function of
the driving frequency in case $h\ll h_p$. For small $\omega$ the area is diminished 
proportional to $\omega$ (exponent $0.98\pm0.01$), as expected.}
\label{fig:hyst_flaeche_hk}
\end{figure}
(a) $h\ll h_p$. In this situation, the hysteresis consists of a single
loop. The outer loops, visible in Fig. \ref{fig:hystereses}, are absent.
We expect the loop area to be given by $A_\text{hyst}\approx v_{h_0}h\sim\omega^\kappa$.
Indeed, our numerical solution shows that the area of the hysteresis vanishes
proportional to $\omega$, independent of the type of disorder
correlator, as shown in Fig. \ref{fig:hyst_flaeche_hk}.

\begin{figure}
\includegraphics[width=\columnwidth]{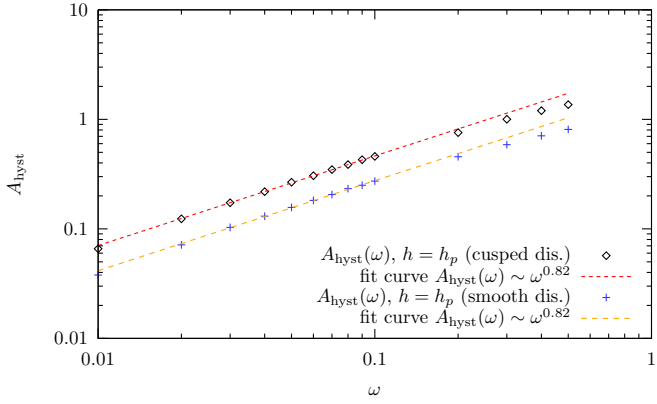}
\caption{The area of the hysteresis loop is plotted as a function of
the driving frequency in case $h\simeq h_p$. For small $\omega$ the area vanishes 
with an exponent $0.82\pm0.01$, independent of the type of disorder.}
\label{fig:hyst_flaeche_hp}
\end{figure}
(b) $h=h_p$. For this case, the hysteresis loop is still single (no double hysteresis)
and the hysteresis area decreases with the frequency as 
$A_\text{hyst}\sim\omega^{0.82}$ (cf. Fig. \ref{fig:hyst_flaeche_hp}), still independent
of the disorder correlator.

\begin{figure}
\includegraphics[width=\columnwidth]{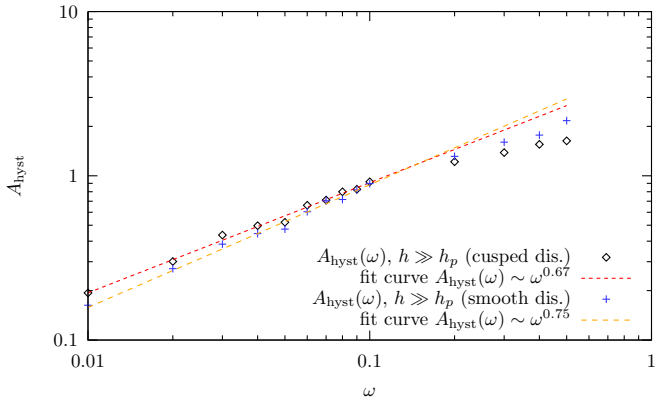}
\caption{The area of the hysteresis loop is plotted as a function of
the driving frequency in case $h\gg h_p$. The diminution of the area
with $\omega$ can be described as a power law with different exponents for
cusped ($\alpha=0.67\pm0.03$) and smooth disorder ($\alpha=0.75\pm0.04$).}
\label{fig:hyst_flaeche_hg}
\end{figure}
(c) $h\gg h_p$. Now, we face the situation of a double hysteresis and moreover,
the behaviour of the hysteresis area as $\omega\to 0$ now depends
on the shape of the disorder correlator. We find $A_\text{hyst}\sim\omega^\alpha$ 
with $\alpha\simeq 0.67\approx 2/3$ for cusped and $\alpha\simeq 0.75=3/4$ 
for smooth disorder. This is shown in
figure \ref{fig:hyst_flaeche_hg}.

So far, our results suggest that the scaling exponents are insensitive to the
nature of the disorder correlator as long as the force amplitude does not exceed 
the threshold $h_p$. These findings seem to milden the non-universality 
conclusion by Fisher \cite{Fisher:PRB85},
who considered the response of a charge density wave system to an ac 
force in addition to dc driving.
He distinguished
different distributions of the random amplitude (pinning strength) 
of the disorder potential in addition to a random phase,
and found a strong dependence of the behaviour on the type of disorder
both above and below threshold.

For large frequencies, the area of the hysteresis loop vanishes as well.
Above a certain crossover frequency, which depends on $h$, the motion of
the particle is restricted to one minimum of the potential.
Thus, for large enough frequencies we can approximate the potential by a 
harmonic one, such that the equation of motion for the disorder averaged position
becomes
\begin{align}
v(t)=h\,\cos\omega t - w\,z(t),
\end{align}
which has the solution
\begin{align}
v(t)={w\omega h\over w^2+\omega^2}\left[-\sin\omega t+{\omega\over w}\cos\omega t\right].
\end{align}
Thus, using Eq. (\ref{eq:ahystdef}) we find for the hysteresis loop area
\begin{align}
A_\text{hyst}(\omega)&=\int\limits_0^Tv(t)\dot h(t)\,dt
={w\omega^2h^2\over w^2+\omega^2}{\pi\over\omega}\sim\omega^{-1},
\end{align}
where the last expression gives the asymptotics for large $\omega$.

The decay of the hysteresis loop area for large and small frequencies of the
driving force requires the existence of a maximum. This maximum is found to be
proportional to the resonance frequency of the typical
disorder potential wells $\omega_r=\gamma u_0/\ell$, 
which equals 1 in our units. The proportionality factor is of order unity, and
is found different for small driving fields (single hysteresis) and large
drivings (double hysteresis). 

\section{Conclusions}

For the mean-field theory of driven elastic manifolds in disordered
systems, we have worked out the scaling behaviour of the velocity as a 
function of the dc-driving force close to depinning by extending
Fisher's \cite{Fisher:PRB85} arguments for charge density waves.
The scaling exponents are found to be different
for disorder with smooth and cusped correlator.
Our analytical findings are supported by a numerical treatment.

Furthermore, we have investigated the small-frequency behaviour of
quantities that characterise the velocity hysteresis in case the system
is exposed to an ac-driving. We found that the frequency scaling exponents of
the remanent velocity $v_{h_0}$ and the velocity at the depinning field $v_{h_p}$
do not depend on the presence or absence of a  cusp-singularity at the origin
of the disorder correlator.
This also holds for the frequency exponents of the
hyteresis loop area as long as the amplitude of the driving does not exceed
the depinning force $h_p$. For force amplitudes above the depinning threshold,
our numerical treatment yields different exponents for smooth and cusped
correlators.

\begin{acknowledgments}
For fruitful discussions we are grateful to G. M. Falco and A. A. Fedorenko.
Moreover, we want to acknowledge financial support by 
Sonderforschungsbereich 608.
\end{acknowledgments}

%
%

\appendix

\section{\label{app:realisation}%
Applicability of the mean field theory%
}

Mean field theory is valid above the upper critical interface 
dimension $D_c$,
where the interface is not rough and there is no depinning transition
for weak disorder.
Below this critical dimensionality, 
the interface is soft enough to adapt to  the random potential. 
Hence, it becomes rough and
is pinned for small driving forces.
An FRG analysis of the depinning transition shows, that the exponents 
become
non-classical and the disorder correlator develops a cusp singularity
\cite{NSTL:JP2F}. The exponents can be expressed in terms
of an $\epsilon$-expansion, where $\epsilon=D_c-D$, and the
zeroth order $O(\epsilon^0)$ is given by the critical exponents of
mean-field theory with a cusped disorder correlator
\cite{NarayanFisher:PRB93}.

For systems with short range interaction it turns out that $D_c=4$,
so interface dimensionalities above $D_c$ seem to be a purely academic 
problem.
There are, however, systems with long range forces for which the upper
critical dimension $D_c$ is reduced. 
For example, for interfaces in systems with long range 
dipolar interaction it has been argued in Ref. \cite{Natter:JPA88} 
that the upper critical dimension decreases to $D_c=3$.
A further decrease of the critical dimension for the statics of
interfaces is achieved for domain walls in ferroelectric materials with a
piezo effect in the paraphase. An example for such a material is KDP.
A peculiarity of such systems is that interfaces between different
ferroelectric domains are allowed to be oriented
along some distinguished orientations in the crystal only.
This is because ferroelectric domain walls are at the same time ferroelastic 
domain walls which have to fulfill certain mechanical compatibility 
relations \cite{Fousek:JAP68,Sapriel:PRB75}.
Looking at the statics, one finds that in the presence of random field disorder 
the interface is not rough for $D\ge 2$
\cite{Schuetze:phd,CDW:PRE04}.
In the following, we are going to take a look on the impact
of these long range forces on the dynamical properties.

Firstly, we note that our model equation of motion (\ref{eq:mean_field0})
can be written in a more general fashion as
\begin{align}
\label{eq:eomgen}
{1\over\gamma}\partial_t z(\vec x,t)=&
u_0g(\vec x,z)+h(t)\nonumber\\
&+\int d^d\vec x'\,\Gamma_\text{el}(\vec x-\vec x')z(\vec x',t),
\end{align}
where 
\begin{align}
\Gamma_\text{el}(\vec x)=
\int_{\vec k}e^{i\vec k\vec x}\Gamma_\text{el}(\vec k).
\end{align}
In (\ref{eq:mean_field0}) we assumed that only short range forces
are present and hence
\begin{align}
\label{eq:h_elsrint}
\Gamma_\text{el}(\vec k)=\Gamma_0k^2.
\end{align}
For the systems with long range forces, discussed above, the functional form 
of $\Gamma_\text{el}(\vec k)$ is different from (\ref{eq:h_elsrint}). 
In systems with dipolar interactions, one finds \cite{Nattermann:JPC83}
\begin{align}
\Gamma_\text{el}(\vec k)=\Gamma_0 k^2+\lambda {k_a^2\over d-2}\left[(k\xi_0)^{2-d}-1\right],
\end{align}
where $\lambda$ measures the strength of the dipolar interaction and $\xi_0$ is of
the order of the lattice spacing. 
Here, $k_a$ is the component
of $\vec k$ along the orientation of the dipoles. 
A further increase of the domain wall stiffness is achieved for domain walls 
in ferroelectric materials with a piezo effect in the paraphase. 
Integrating out the elastic degrees of freedom in such a (3d) system
leads to an interface Hamiltonian of the form \cite{Schuetze:phd}
\begin{align}
\label{eq:h_elpiezoint}
\Gamma_\text{el}(\vec k)=\Gamma_0 k^2+\lambda{k_a^2\over k}+ck.
\end{align}
Here, $c$ is a constant that depends on the strength of the piezo coupling
as well as on the elastic constants and $\lambda$ depends on the strength of the 
dipolar coupling, the piezo coupling coefficient and the elastic constants. 
Moreover, $k_a$ denotes the component of $\vec k$ in the direction of the 
polarisation.

In the following, we focus on the case of a dc-driving field $h(t)\equiv h$.
For the perturbative determination of the velocity, it turns out to
be useful to go over to a co-moving
frame $z(\vec x,t)=vt+\zeta(\vec x,t)$.
Expanding the correction $\zeta(\vec x,t)$ in a power series in $u_0$,
\begin{align}
\zeta(\vec x,t)=\sum\limits_{n=0}^\infty u_0^n\zeta_n(\vec x,t),
\end{align}
one can solve Eq. (\ref{eq:eomgen}) perturbatively by comparing
equal orders of $u_0$. Following the computation
in Ref. \cite{NSTL:APL} (which is similar to that presented in App. 
\ref{app:pert} below), we obtain a self-consistent perturbative 
expression for $v$ 
(the following equation corresponds to Eq. (29) in Ref. \cite{NSTL:APL})
\begin{align}
\label{eq:pert_exp_for_v}
v-\gamma h&\simeq\int_q\int_{\vec k}{i u_0^2q\Delta_q\over 
\Gamma_\text{el}(\vec k)-i{vq\over\gamma}}\nonumber\\
&=\int_q\int_{\vec k}{iq\Delta_qu_0^2(\Gamma_\text{el}(\vec k)+i{vq\over\gamma})
\over[\Gamma_\text{el}(\vec k)]^2+{v^2q^2\over\gamma^2}}.
\end{align}
Here, $\Delta_q$ denotes the Fourier transform of the disorder
correlator defined in Eq. (\ref{eq:diskorrdef}).
For $v\to 0$, our choice (\ref{eq:h_elsrint}) for $\Gamma_\text{el}(\vec k)$ 
leads to a divergence of the integral in (\ref{eq:pert_exp_for_v}) 
when $D<4$. 
The divergence stems from the contribution of small $\vec k$.
In contrast, for a model the elastic interaction of which assumes
the form given by Eq. (\ref{eq:h_elpiezoint}), 
the integral in Eq. (\ref{eq:pert_exp_for_v}) is convergent as long as $D>2$ 
and exhibits only a logarithmic divergence in $D=2$. Thus, in such a system
mean field results may apply for two-dimensional domain walls
with small logarithmic corrections.

\section{\label{app:num}%
Numerics%
}

In this appendix, the details of our numerical methods shall
be explained briefly.
Numerically, we have solved the equation of motion (\ref{eq:eom}) for 20000
different disorder configurations using the well established 
classical Euler method. Before embarking on the details of the
procedure in the two cases of dc and ac driving, respectively, we
want to explain how the different realisations of the disorder forces
are achieved.
\begin{figure}
\begin{center}
\includegraphics[width=\columnwidth]{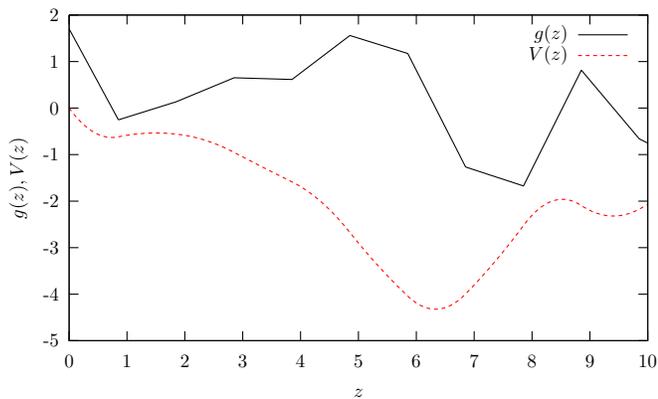}
\caption{An example for a randomly generated disorder realisation
with the method that leads to a statistics with a smooth correlator.
Besides the force $g(z)$, also the disorder potential, determined from
$g(z)=-V'(z)$ is visualised. The potential has a smooth shape.}
\label{fig:glattbsp}
\end{center}
\end{figure}

The random force associated to the smooth
disorder correlator is generated by concatenating straight lines
at distance 1. The values of the disorder force at the concatenation 
points are randomly chosen out of a bounded
interval $[-m;m]$. 
The bounds $m$ for this interval are determined such that the correlator
fulfills $\Delta(0)=1$. We found $m\simeq 2.1$. 
Moreover, the position $s$ of the concatenation point closest
to the origin $z=0$ has been determined randomly out of the interval
$[-{1\over 2};{1\over 2}]$. This method has already been used earlier
by A. Glatz \cite{Glatz:phd}. A sample configuration is depicted in Fig.
\ref{fig:glattbsp}.
\begin{figure}
\begin{center}
\includegraphics[width=\columnwidth]{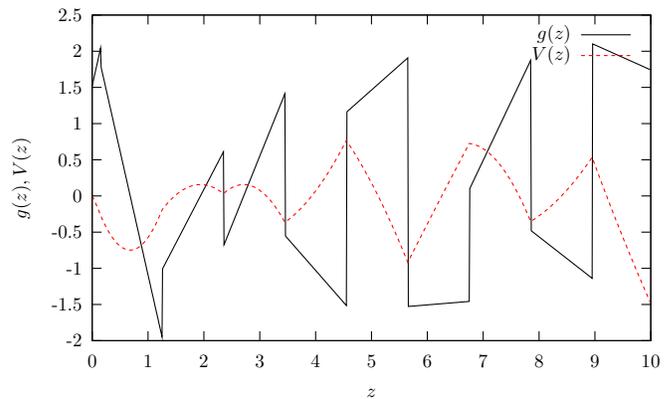}
\caption{An example for a randomly generated disorder realisation
with the method that leads to a statistics with a cusped correlator.
Besides the force $g(z)$, also the disorder potential, determined from
$g(z)=-V'(z)$ is visualised. It can be seen that most of the 
extrema of the potential reveal a cusp singularity.}
\label{fig:cuspbsp}
\end{center}
\end{figure}

To realise disorder forces with Gaussian statistics that are correlated with
a cusp singularity at the origin, we have also used straight lines, the
extension in $z$-direction being again 1. In contrast to the former case,
these lines are now concatenated discontinuously. The values of
$g(z)$ at both endpoints of a segment are
determined randomly from a bounded interval $[-m;m]$. 
The size of this interval is again determined in such a way that the 
disorder correlator obeys $\Delta(0)=1$ with the same result of $m\simeq 2.1$ 
as before.
Also the position $s\in[-{1\over 2};{1\over 2}]$
of the jump closest to $z=0$ is created randomly.
An example for such a configuration is shown in Fig. \ref{fig:cuspbsp}.
In App. \ref{app:cuspcorr} we discuss the latter generation technique
in more detail and also derive the force correlator.

\begin{figure}
\begin{center}
\includegraphics[width=\columnwidth]{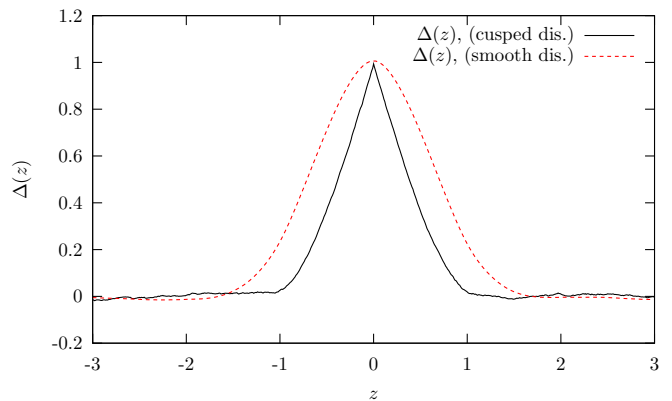}
\caption{The numerically determined correlators
for the disorder forces generated by the two different 
methods described in the text. Either disorder correlator
decays on a length scale of order unity and obeys $\Delta(0)=1$.}
\label{fig:numdiskorr}
\end{center}
\end{figure}

Moreover, we have verified numerically, that the assumed shapes for the
disorder correlators are reflected by our two generation techniques. The result
is shown in Fig. \ref{fig:numdiskorr}. As required, one correlator is
smooth and the other shows a cusp singularity at the origin, perfectly
in agreement with the analytic result (cf. Eq. (\ref{eq:dcorrcuspanal}) below). 
In both
cases, the correlations decay to zero on a length scale $\ell\simeq 1$.

To solve the equation of motion with a constant driving force for several
values of $\Gamma$ and $h$, we have chosen the initial condition $v(0)=0$ and
$z(0)=0$. The time steps are chosen to be of size $2^{-10}$ and we have
simulated the equation of motion for $2^{14}$ such time steps. To avoid that
our results are spoiled by transience effects, we have only taken into
account the values of $v$ for the last $2^{10}$ time steps.

Our results that we have obtained for the ac driving case rely on the
initial conditions $z(0)=0$, $v(0)=h$ with $h(t)=h\,\cos\omega t$.
Before making any measurements we have been waiting for at least 2 periods
for transience effects to diminish. The values for $v(h=0)$ and $v(h=h_p)$
are measured then over 3 periods with 2 datapoints per period.

In all cases where we have numerically determined important exponents,
we have given error estimates. These include the statistical deviations.

\section{\label{app:pert}%
Perturbation theory%
}

To set up the perturbation expansion for our equation of motion
(\ref{eq:dceom})
\begin{align}
\partial_tz=\Gamma(\rav{z}-z)+h+g(z),
\end{align}
we write $z(t)=\rav{z(t)}+\zeta(t)$, $\partial_tz=v+\partial_t\zeta$,
so that $\rav{\zeta}=0$. This yields two coupled differential equations
\begin{align}
\label{eq:gl_fuer_v}
v&=h+\rav{g(\rav{z}+\zeta)}
\\
\label{eq:recursion_for_zeta}
(\partial_t+\Gamma)\zeta&=g(\rav{z}+\zeta)-\rav{g(\rav{z}+\zeta)}.
\end{align}
The differential operator on the left hand side of 
(\ref{eq:recursion_for_zeta}) has the fundamental solution
\begin{align}
(\partial_t+\Gamma)G(t)&=\delta(t),\\
G(t)&=\Theta(t)\,e^{-\Gamma t}.
\end{align}
Now, for large enough $\Gamma$, we can expect $\zeta(t)\ll 1$ and perform
a Taylor expansion of the disorder force
\begin{align}
\label{eq:taylor_dis}
g(\rav{z}+\zeta)&=\sum\limits_{n=0}^\infty{g^{(n)}(\rav{z})\over n!}\zeta^n.
\end{align}
Thus, (\ref{eq:gl_fuer_v}) gives us
\begin{align}
\label{eq:pert_exp_v}
v&=h+\rav{g(\rav{z})+g'(\rav{z})\zeta+\ldots}
\nonumber\\
&=h+\rav{g'(\rav{z})\zeta(t)+\ldots}
\end{align}
For $\zeta(t)$, we use the lowest order result from the iteration
(\ref{eq:recursion_for_zeta}) combined with (\ref{eq:taylor_dis})
\begin{align}
\zeta(t)&=\int\limits_{t_0}^\infty dt'\,G(t-t')\left[
g(\rav{z(t')}+\zeta)-\rav{g(\rav{z(t')}+\zeta)}\right]
\nonumber\\
&=\int\limits_{t_0}^t dt'\,e^{-\Gamma(t-t')}\left[g(\rav{z(t')})+\ldots\right],
\end{align}
where $t_0$ is the time at which we fix our initial value problem.
Sending $t_0\to-\infty$, we obtain
\begin{align}
v&=
h+\int\limits_{-\infty}^tdt'\,e^{-\Gamma(t-t')}\rav{g'(\rav{z(t)})g(\rav{z(t')})}+\ldots
\nonumber\\
&=h+\int\limits_{-\infty}^tdt'\,e^{-\Gamma(t-t')}\Delta'[\rav{z(t)}-\rav{z(t')}].
\end{align}
A change of the integration variable $t'\to t-t'$ and noting, that
$\rav{z(t)}=vt$ then immediately yields the
result given by (\ref{eq:vpertdc}).

\section{\label{app:cuspcorr}%
Generation of disorder forces with a cusped correlator%
}

In this appendix, we propose a generation technique for disorder
forces with a cusp singularity in the correlator and prove, that the
cusp is indeed present.
The method described here has also been employed in our numerical
analysis, cf. App. \ref{app:num}.

Consider a function $g(z)$, that is constructed as follows.
We decompose the $z$-axis in intervals $I_i=[z_i;z_{i+1}]$ of length $1$. 
The starting
point of each interval $I_i$ is given by
\begin{align}
z_i=i+s,
\end{align}
where $i$ is an integer
(the label for the interval) and $s$ is a random number, uniformly
distributed in the interval $[-{1\over 2};{1\over 2}]$.
The function $g(z)$ shall now be given piecewise for each
interval $I_i$ as a straight line.
This line is determined by the left boundary point $(z_i,\alpha_i)$
and the right boundary point $(z_{i+1},\beta_i)$, where
$\alpha_i$ and $\beta_i$ are chosen randomly out of a bounded
interval $[-m;m]$. The value of $m$ will be determined further down
in such a way, that $\Delta(0)=1$. 

In more mathematical terms, a specific realisation of the
function $g(z)$ is specified by the random number $s\in[-{1\over 2}:{1\over 2}]$
and two stets of boundary values $\{\alpha_i\}$ and $\{\beta_i\}$.
Using the indicator function $K(z;z_i,z_{i+1})$ for each interval $I_i$,
defined by
\begin{align}
K(z;z_i,z_{i+1})
=\left\lbrace
\begin{matrix}1&\quad z_i\le z< z_{i+1}\\ 0&\quad\text{otherwise}
\end{matrix}\right.,
\end{align}
the function $g(z)$ is explicitly given by
\begin{align}
\label{eq:forcefunction}
g(z,[s;\{\alpha_i\};&\{\beta_i\}])=\sum_iK(z;z_i,z_{i+1})
\nonumber\\&\times
\big[(\beta_i-\alpha_i)(z-i-s)+\alpha_i\big].
\end{align}
\begin{figure}
\begin{center}
\includegraphics[width=.8\columnwidth]{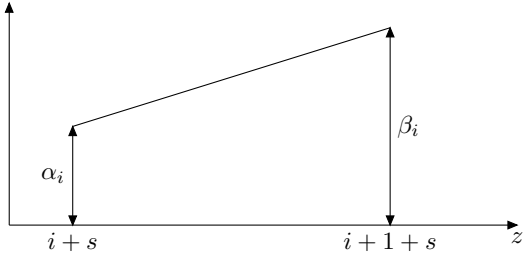}
\caption{A segment of the disorder force with a cusp singularity
in the correlator. The values $\alpha_i$ and $\beta_i$ are
chosen out of a bounded interval.
}
\label{fig:disillu}
\end{center}
\end{figure}%
Examples for typical configurations of $g(z)$ are depicted in the
Figs. \ref{fig:intersection2} and \ref{fig:cuspbsp}. An illustration
for a single segment is sketched in Fig. \ref{fig:disillu}.
In the following, we are going to show, that disorder
forces $g(z)$ given by (\ref{eq:forcefunction})
fulfill our requirements $\av{g(z)}=0$ and
$\av{g(z)g(z')}=\Delta(z-z')$, where $\Delta(z-z')$ obeys
$\Delta(0)=1$, decays to zero over a length scale of order 1
and shows a cusp singularity at the origin.
Straightforwardly, we find
\begin{align}
\rav{g(z)}=\int\limits_{-{1\over 2}}^{1\over 2}ds
\int\limits_{-m}^m{d\beta d\alpha\over 4m^2}\>
[(\beta-\alpha)(z-i-s)+\alpha]=0.
\end{align}
To calculate the second moment $\av{g(z)g(z')}$, 
we have to distinguish two cases.
If $|z-z'|\ge 1$, the points $z$ and $z'$ have to belong to two
different intervals $z\in I_k$, $z'\in I_{k'}$, because their distance
is then larger than the size of an interval and hence
\begin{align}
g(z)g(z')=&\sum\limits_{i,j}K(z;z_i,z_{i+1})\big[(\beta_i-\alpha_i)(z-z_i)+\alpha_i\big]
\nn\\&\times
K(z';z_j,z_{j+1})\big[(\beta_j-\alpha_j)(z'-z_j)+\alpha_j\big]
\nn\\
=&\big[(\beta_k-\alpha_k)(z-k-s)+\alpha_k\big]
\nn\\&\times
\big[(\beta_{k'}-\alpha_{k'})(z'-{k'}-s)+\alpha_{k'}\big].
\end{align}
This gives for the correlator
\begin{align}
\rav{g(z)g(z')}=&\int\limits_{-{1\over 2}}^{1\over 2}ds\int\limits_{-m}^m
{d\beta_k d\beta_{k'}d\alpha_k d\alpha_{k'}\over 16m^4}
\nn\\
&\times\,[(\beta_k-\alpha_k)(z-z_k)+\alpha_k]
\nn\\
&\times\,[(\beta_{k'}-\alpha_{k'})(z'-z_{k'})+\alpha_{k'}]
\nn\\
=&0.
\end{align}
On the other hand, in case $|z-z'|<1$, those realisations which do not have
a jump in between $z$ and $z'$, i.e. for which $z$ and $z'$
belong to the same interval $I_k$, give a finite contribution. These correspond to
values of $s$, for which 
\begin{align}
\zeta\equiv |z-z'|+\min(z,z')-i-s
\end{align}
obeys $|z-z'|<\zeta<1$.
Instead of integrating over $s$, it is easier to integrate over $\zeta$ 
which gives us
\begin{align}
\rav{g(z)g(z')}=&\int\limits_{|z-z'|}^1\!\!d\zeta
\int\limits_{-m}^m{d\beta_k d\alpha_k\over 4m^2}\>
[(\beta_k-\alpha_k)\zeta+\alpha_k]
\nn\\
&\times\,
[(\beta_k-\alpha_k)(\zeta-|z-z'|)+\alpha_k]
\nn\\
\label{eq:dcorrcuspanal}
=&{2m^2\over 9}\left[1-{3\over 2}|z-z'|+{1\over 2}|z-z'|^3\right].
\end{align}
As required, the correlator exhibits a cusp singularity at the origin and
decays to 0 on a length scale $\ell=1$.
To fulfill $\Delta(0)=1$, we have to take $m=3/\sqrt 2\simeq 2.1$, which
is in agreement with our numerical technique, described in App. \ref{app:num}.

\section{\label{app:energy}%
Derivation of the potential energy balance%
}

The equation of motion (\ref{eq:eom}) can be considered to follow
from a Hamiltonian
\begin{align}
\partial_tz=-{\delta H[z]\over\delta z},
\end{align}
with
\begin{align}
H[z]&=E[z]-h(t)z(t)
\\
E[z]&={\Gamma\over 2}\left(z(t)-\av{z(t)}\right)^2+V(z).
\end{align}
Here, $E[z]$ denotes the total potential energy, $V(z)$ is the disorder potential
related to $g(z)$ via $g(z)=-V'(z)$. The change of the potential energy
in time thus follows as
\begin{align}
\partial_tE[z]&=\Gamma(z-\av{z})(\partial_tz-v(t))-g(z)\partial_tz
\nonumber\\
&=\Gamma(\av z-z)v(t)-\left[\Gamma(\av z-z)+g(z)\right]\partial_tz.
\end{align}
Using Eq. (\ref{eq:eom}), we can replace the term in the square brackets 
and obtain
\begin{align}
\partial_tE[z]&=\Gamma(\av z-z)v(t)+\left[h(t)-\partial_tz\right]\partial_tz.
\end{align}
Taking the disorder average readily yields the result, stated in Eq. (\ref{eq:potchange}).


\bibliography{wand}

\begin{thebibliography}{38}
\bibitem[{not()}]{note1}
\bibinfo{note}{Here and below we follow closely the arguments of D.S. Fisher
  \cite{Fisher:PRB85} who considered the slightly different charge density wave
  problem.}
\end{thebibliography}

\end{document}